\documentclass[11pt]{article}
\usepackage[margin=1in]{geometry}
\usepackage{graphicx}
\usepackage{tikz}
\usepackage{pgfplots}
\pgfplotsset{compat=1.17}
\usepackage{amsmath}
\usepackage{booktabs}
\usepackage{hyperref}
\hypersetup{colorlinks=true, linkcolor=blue, citecolor=blue, urlcolor=blue}

\title{Mixed Reality Guidance of a Surgical Scalpel Using Magic Leap:\\ Evaluation on a 3D-Printed Liver Phantom}
\author{%
    Alice Y. \thanks{Department of Computer Science. \texttt{}} \and 
    Michael B. \thanks{Department of Hepatobiliary Surgery. \texttt{}} \and 
    Catherine T.\thanks{Biomedical Engineering Institute, MedTech Labs. \texttt{catherine.inn@medtechlabs.com}} \and 
    Hu Guo \thanks{University of Cincinnati. \texttt{goudn@mail.uc.edu}}
}

\begin{document}
\maketitle

\begin{abstract}
\noindent \textbf{Abstract:} Augmented and mixed reality (MR) systems have the potential to improve surgical precision by overlaying digital guidance directly onto the operative field. This paper presents a novel MR guidance system using the Magic Leap head-mounted display to assist surgeons in executing precise scalpel movements during liver surgery. The system projects holographic cues onto a patient-specific 3D-printed liver phantom, guiding resection along a predetermined path. We describe the system design, including preoperative modeling, registration of virtual content to the phantom, and real-time visualization through the Magic Leap device. In a controlled phantom study, surgical trainees performed resection tasks with and without MR guidance. Quantitative results demonstrated that MR guidance improved cutting accuracy (mean deviation from planned path was reduced from 5.0~mm without AR to 2.0~mm with AR guidance) and efficiency (mean task time decreased from 55~s to 32~s). These improvements of approximately 60\% in accuracy and 40\% in speed underscore the potential benefit of MR in surgical navigation. Participants reported that the Magic Leap visualization enhanced depth perception and confidence in locating tumor boundaries. This work provides a comprehensive evaluation of an MR-assisted surgical guidance approach, highlighting its feasibility on a realistic organ phantom. We discuss the technical challenges (registration accuracy, line-of-sight, user ergonomics) and outline future steps toward clinical translation. The results suggest that Magic Leap-based MR guidance can significantly augment a surgeon’s performance in delicate resection tasks, paving the way for safer and more precise liver surgery.
\end{abstract}

\section{Introduction}
Intraoperative guidance technologies play a critical role in enhancing surgical precision and safety. Augmented reality (AR) and mixed reality (MR) interfaces, which overlay computer-generated images onto the surgeon’s view of the real world, have attracted growing interest as tools for surgical navigation \cite{Doughty2022, Zari2023, wang2025mruct}. By allowing surgeons to visualize anatomical structures, tumor boundaries, or planned incision lines within the actual operative field, AR/MR systems can reduce reliance on mental mapping of radiologic images and improve hand–eye coordination during complex procedures \cite{Groves2022}. Recent advances in optical see-through head-mounted displays (OST-HMDs) such as the Microsoft HoloLens and Magic Leap have further spurred development of AR applications in medicine \cite{Doughty2022, Zari2023}. These wearable devices enable surgeons to maintain an egocentric view of 3D guidance cues without diverting gaze to external monitors, potentially streamlining workflow in the operating room.

Liver surgery is a domain that could particularly benefit from AR/MR guidance. Precise navigation is required to resect tumors while preserving critical vasculature and adequate margins. Conventional image guidance methods (e.g., ultrasound or neuronavigation-like pointer systems) provide valuable information but often require the surgeon to interpret this information on separate screens or through probes, which can be mentally demanding and error-prone. Three-dimensional printed surgical guides have also been explored for liver resection planning \cite{Ribeiro2024}, but such guides are static and cannot adapt intraoperatively. In contrast, a mixed reality approach can dynamically display patient-specific anatomical information, such as the tumor location or safe cutting planes, directly on the organ in situ. Prior studies have demonstrated the promise of AR in open and laparoscopic liver surgery \cite{Ribeiro2024}, as well as other surgical fields. For example, AR has been used to overlay hepatic vasculature CT models onto patients for laparoscopic resection guidance \cite{Ribeiro2024}, and similar OST-HMD systems have been applied in orthopedic, neurologic, and endovascular procedures \cite{Uhl2022, Frisk2022, Schneider2021}. However, there remain significant challenges in achieving the accuracy, robustness, and user acceptance required for clinical adoption of AR in the operating room \cite{Doughty2022}. Reported system accuracy in phantom models is on the order of 2–5~mm, which is promising, but human factors such as depth perception, ergonomic comfort, and occlusion of real-world cues must be carefully addressed before AR can be widely adopted in surgical practice \cite{Doughty2022}.

The Magic Leap head-mounted device represents a state-of-the-art OST-HMD that has recently gained interest for surgical applications. It features a wide field of view and high resolution display with stereoscopic capabilities, which are crucial for visualizing complex 3D anatomy \cite{Zari2023}. A user study comparing Magic Leap 1 to HoloLens 2 found that surgeons preferred the Magic Leap for its superior visual clarity and ease of interaction \cite{Zari2023}. These attributes suggest the Magic Leap could serve as an effective platform for intraoperative MR guidance. Indeed, preliminary investigations have used Magic Leap in surgical contexts: Uhl \textit{et al.} employed Magic Leap to assist femoral artery punctures on a vascular phantom, achieving median targeting errors around 1~mm \cite{Uhl2022}, and Frisk \textit{et al.} demonstrated pedicle screw placement in spine phantoms with Magic Leap-based AR navigation, reporting high placement accuracy \cite{Frisk2022}. Building on this foundation, we aim to explore the utility of Magic Leap for guiding freehand surgical resections.

In this paper, we present an MR guidance system that projects interactive holographic cues via the Magic Leap headset to guide a surgeon’s scalpel during liver resection. To our knowledge, this is the first work to integrate Magic Leap for guiding a cutting tool along a specified trajectory on a 3D organ model. We focus on a clinically inspired scenario of liver tumor resection. A life-sized liver phantom with embedded tumor is used to simulate surgery in a risk-free environment while providing realistic tactile feedback. The MR system is used to overlay the planned resection line and internal tumor location onto the phantom, aiding the surgeon in executing the cut with high precision. We evaluate the system in a comparative study where users perform resection tasks on the phantom with and without MR assistance. Key outcomes include the geometric accuracy of the excised area relative to the preoperative plan and the time efficiency of the task. We also gather qualitative feedback from participants regarding the usability and perceived utility of the MR guidance.

The contributions of this work are threefold: (1) We develop a novel MR guidance technique using Magic Leap to project surgical cut trajectories onto patient-specific anatomy, including methods for calibration and registration on a 3D-printed phantom. (2) We provide a thorough experimental evaluation of the system’s performance, demonstrating significant improvements in cutting accuracy and speed under MR guidance. (3) We discuss the practical considerations and limitations encountered, offering insights for future developments in mixed reality surgical navigation. Ultimately, our results indicate that MR guidance with Magic Leap can enhance a surgeon’s spatial awareness and execution in a controlled experimental setting, supporting the case for further investigations toward intraoperative use on patients.

\section{Related Work}
\subsection{Augmented Reality in Surgery}
Augmented reality has been an area of active research in surgical navigation for over a decade. Early implementations often relied on external displays or projection systems to overlay images onto the operative field. With the advent of modern OST-HMDs like the HoloLens and Magic Leap, there has been a surge in head-mounted AR solutions across various surgical specialties \cite{Doughty2022}. Doughty \textit{et al.} \cite{Doughty2022} conducted a systematic review of OST-HMD use in surgery and noted that while technical feasibility has been demonstrated in numerous pilot studies, challenges such as tracking accuracy, limited field of view, and user comfort have constrained widespread adoption. They reported that median targeting errors of 2--5~mm are common in phantom and cadaver studies using AR HMDs \cite{Doughty2022}, which is encouraging for many surgical tasks (such as needle placement) but improvements are needed for high-precision requirements (e.g., neurosurgery).

Several notable applications of AR in surgery have been reported in literature. In neurosurgery, AR has been used to assist in cranial tumor resection and catheter placement. For example, Schneider \textit{et al.} \cite{Schneider2021} developed an AR-assisted ventriculostomy system using an OST-HMD to display the planned trajectory of a catheter in the patient’s brain, enabling accurate placement of external ventricular drains. Their system superimposed a holographic trajectory line onto the patient and achieved successful insertions with minimal deviation from the ideal path \cite{Schneider2021}. Similarly, Ivan \textit{et al.} explored AR for incision planning in cranial surgery, finding that surgeons could place skin incisions more precisely when guided by AR projections of underlying anatomy \cite{Schneider2021}. In orthopedic spine surgery, AR navigation has been applied to pedicle screw insertion. Frisk \textit{et al.} \cite{Frisk2022} evaluated the Magic Leap headset for this purpose, overlaying 3D trajectories for screw placement onto spine phantom models. They reported that all screws were accurately placed with the AR system, with placement errors comparable to conventional navigation systems \cite{Frisk2022}. Other groups have used the HoloLens for similar pedicle screw guidance tasks, demonstrating the general feasibility of AR in reducing localization errors in rigid anatomy \cite{Frisk2022}.

In the domain of interventional radiology and endovascular procedures, AR has shown promise for guiding instruments to targets that are otherwise invisible to the naked eye. Uhl \textit{et al.} \cite{Uhl2022} presented a mixed reality system combining Magic Leap with an optical tracking platform to guide femoral artery punctures. Radiopaque fiducials on a vascular phantom were registered to their corresponding points in a CTA-derived 3D model, which was then visualized in situ through the headset. This allowed the operator to see a virtual vessel tree inside the phantom and the ideal needle entry point on the surface. They achieved a median axial puncture error of only 1.0~mm and sagittal error of 1.1~mm, indicating highly accurate alignment \cite{Uhl2022}. Similarly, in urology, Porpiglia \textit{et al.} \cite{Porpiglia2022} used an AR hologram (via HoloLens) to guide percutaneous kidney punctures during nephrolithotomy. Their approach involved preoperative planning of an access path on CT, followed by intraoperative AR overlay of that path onto the patient’s anatomy. The AR-guided puncture was successful in all cases, illustrating AR’s potential to improve accuracy in procedures where line-of-sight to the target is obstructed \cite{Porpiglia2022}.

\subsection{AR in Hepatic Surgery and Training}
Liver surgery has seen growing interest in AR and MR applications, primarily in the context of surgical planning and intraoperative navigation for tumor resections. A number of studies have focused on AR in minimally invasive (laparoscopic) liver resections, where the challenge is to correlate preoperative imaging with the limited view from a laparoscope. Ribeiro \textit{et al.} \cite{Ribeiro2024} conducted a phantom study of AR-guided laparoscopic liver resection. They created a 3D-printed liver phantom with embedded “tumors” and overlayed a virtual 3D model onto the laparoscopic video feed in real time. Their quantitative assessment showed that AR guidance enabled more accurate resection of intraparenchymal tumors, with surgeons achieving closer conformity to the planned resection margins compared to unaided resection \cite{Ribeiro2024}. This suggests AR can convey subsurface tumor locations effectively, even through a minimally invasive approach. Another study by Kitagawa \textit{et al.} \cite{Kitagawa2022} demonstrated the use of an MR wearable (HoloLens-based) for holographic navigation during laparoscopic cholecystectomy. The system displayed key anatomy (e.g., bile ducts and vessels) as holograms during the procedure, helping the surgeon avoid critical structures. Although that study was a pilot on a single case, it showed the feasibility of integrating MR guidance into standard laparoscopic workflows \cite{Kitagawa2022}.

Beyond the operating room, AR and MR are increasingly used in surgical education and training with lifelike simulators. The combination of 3D-printed anatomical models (phantoms) and AR visualization provides trainees with visual cues and real-time feedback, potentially accelerating skill acquisition. For instance, Groves \textit{et al.} \cite{Groves2022} developed a first-person MR guidance system for central venous catheterization training. Using an AR headset, they projected the needle trajectory and underlying venous anatomy onto a manikin. In a comparative study, novices who used the MR guidance achieved higher success rates and improved accuracy of catheter placement versus those using ultrasound alone \cite{Groves2022}. This highlights how MR can complement or enhance traditional guidance modalities (like ultrasound) by intuitively fusing imaging information with the physical task. Our work continues in this vein by applying MR to an open surgical task on a complex organ phantom. The liver phantom scenario tests not only accuracy of overlay but also how the surgeon interacts with holographic guidance while performing a freehand, highly dexterous task (the act of surgical cutting).

\subsection{Magic Leap and Advanced AR Display Technologies}
The Magic Leap One is a lightweight, see-through AR headset equipped with advanced display optics and sensors, and it has drawn attention for medical use due to its high visual fidelity. Zari \textit{et al.} \cite{Zari2023} provided a head-to-head comparison between Magic Leap 1 and Microsoft HoloLens 2 for medical 3D visualization tasks. In their user study, surgeons viewed patient-specific 3D anatomical models (e.g., organ reconstructions from CT scans) on both devices. There were no significant differences in objective performance metrics like frame rate or task completion times; however, subjective feedback favored Magic Leap 1, with users citing better image quality and more natural interaction via its handheld controller \cite{Zari2023}. This suggests that for applications requiring careful inspection of 3D medical imagery, Magic Leap’s display might offer an advantage. It is worth noting that current OST-HMDs including Magic Leap have a fixed focal plane and limited accommodative range, which can cause some difficulty in focusing on very close objects (within 50~cm) \cite{Zari2023}. This limitation is being addressed in newer models (the Magic Leap 2 introduces an autofocus capability and larger field of view), but it is a factor to consider in surgical settings where the working distance is often short. In our study, users wore the Magic Leap at roughly 50--60~cm from the phantom, which is within a comfortable viewing range for the device.

In summary, existing literature demonstrates that AR/MR technology is making inroads in surgery, from guiding needle-based interventions to aiding complex resections. The Magic Leap headset has emerged as a viable platform in this domain, but thus far it has predominantly been studied in controlled settings (phantoms, cadavers) and for tasks like needle insertion or screw placement. There is a gap in exploring its use for guiding continuous surgical actions such as cutting along an arbitrary path. Our work addresses this gap by leveraging Magic Leap for an open surgery scenario on a soft tissue phantom and rigorously evaluating its impact on performance. The following sections describe our MR guidance system design and the experimental methods and results of our phantom study.

\section{Methods}
\subsection{System Overview}
\begin{figure}[tb]
    \centering
    \begin{tikzpicture}[>=stealth, node distance=1.8cm]
        \node[draw, rectangle, fill=blue!10, thick, minimum width=3.5cm, minimum height=1cm, align=center] (model) {3D Liver Model\\(with tumor)};
        \node[draw, rectangle, fill=green!10, thick, minimum width=3.2cm, minimum height=1cm, below of=model, yshift=-1.2cm, align=center] (planning) {Resection Plan\\(cutting path)};
        \node[draw, rectangle, fill=orange!10, thick, minimum width=3.5cm, minimum height=1cm, below of=planning, yshift=-1.2cm, align=center] (app) {Magic Leap App\\(Rendering Engine)};
        \node[draw, shape border rotate=90, fill=gray!20, thick, minimum height=1.2cm, minimum width=1.2cm, below left of=app, xshift=-1.8cm, yshift=-1.0cm, label=below:{\small Tracking Camera}] (tracking) {};
        \node[draw, rectangle, fill=gray!20, thick, minimum width=3.2cm, minimum height=1cm, below right of=app, xshift=1.8cm, yshift=-1.0cm, align=center] (phantom) {3D-Printed Liver Phantom};
        \node[draw, rectangle, fill=yellow!10, thick, minimum width=2.8cm, minimum height=0.8cm, above right of=phantom, xshift=1.0cm, yshift=0.2cm, align=center, font=\small] (surgeon) {Surgeon with Magic Leap HMD};
        \draw[->, thick] (model) -- node[right, font=\small]{plan overlay} (planning);
        \draw[->, thick] (planning) -- node[right, font=\small]{load plan \& model} (app);
        \draw[->, thick] (phantom) -- node[above, sloped, font=\small]{optical markers} (tracking);
        \draw[->, thick] (tracking) -- node[below, sloped, font=\small]{phantom pose} (app);
    
        \draw[->, thick] (phantom) -- node[below, font=\small]{} (surgeon);
        \node[above of=model, font=\small, yshift=0.0cm] {Preoperative};
        \node[right of=app, font=\small, xshift=1.4cm] {Intraoperative};
    \end{tikzpicture}
    \caption{System architecture for Magic Leap-based mixed reality (MR) guidance. A patient-specific 3D liver model with tumor is used to plan the resection path preoperatively. The model and planned cut path are loaded into a Magic Leap application. During surgery, an optical tracking system (or Magic Leap’s spatial mapping) registers the physical 3D-printed liver phantom to the virtual model. The Magic Leap headset then projects the holographic model and cutting guide onto the surgeon’s view of the phantom in real time, allowing the surgeon to follow the planned path with the scalpel.}
    \label{fig:system}
\end{figure}
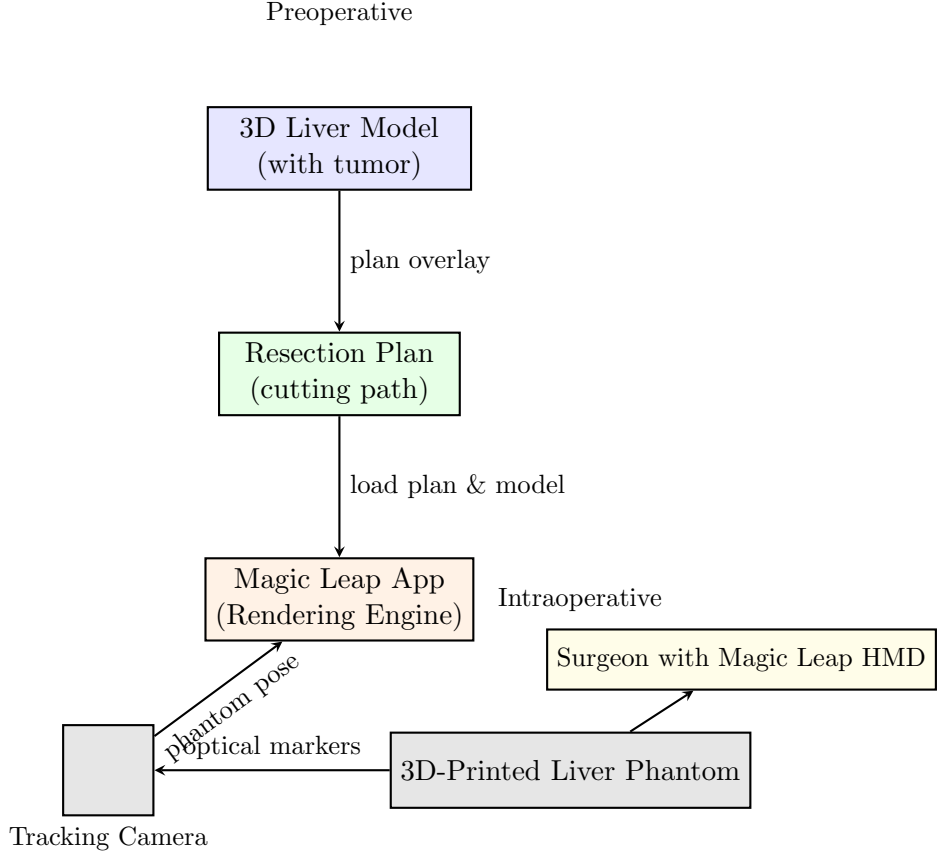

Our mixed reality guidance system is designed to assist a surgeon in performing a liver resection by providing visual overlays through a Magic Leap One head-mounted display. The overall workflow consists of: (1) preoperative modeling and planning, (2) system setup and calibration, and (3) intraoperative guidance visualization. Figure~\ref{fig:system} provides a schematic overview of the system components and data flow.

\paragraph{Magic Leap Headset and Software:} We chose the Magic Leap One as the visualization device due to its high-resolution stereoscopic display and comfortable untethered design. The headset includes outward-facing cameras and sensors that enable spatial mapping of the environment. A custom Unity-based application was developed for the Magic Leap to render the virtual content (the liver model, tumor, and guidance cues) and to handle user interactions. The Magic Leap’s handheld controller was used to allow the surgeon to perform simple interactions, such as toggling the visibility of certain holograms or initiating the registration procedure. The application ran on the Magic Leap’s onboard computer, achieving interactive frame rates (~60~fps) to ensure smooth and stable visualization. The virtual liver model is rendered with partial transparency to allow the user to see both the organ surface and the internal tumor target. The planned resection line is rendered as a bold red curve on the liver surface, clearly distinguishable against the phantom (which is a pale color). To enhance depth perception, we enabled Magic Leap’s default stereoscopic rendering and also provided optional “clip plane” views that could be toggled to see cross-sections of the model, though the latter was not frequently needed during our tasks.

\paragraph{3D Liver Phantom and Tumor Model:} We created a life-sized liver phantom by 3D printing based on a segmented CT scan of an actual human liver with a single tumor. The liver model was printed using a soft resin material to approximate the consistency of liver tissue, and its dimensions (~20~cm across) match a real liver lobe. The phantom includes an embedded mock tumor (a 2~cm spherical inclusion of contrasting color/hardness) situated a few centimeters below the surface on the anterolateral aspect. This provided a realistic target for resection—surgeons would aim to excise the tumor with an adequate margin. The surface of the phantom was painted with a matte, uniform color to facilitate optical tracking (shiny surfaces can cause tracking glare). We also incorporated three small fiducial markers on the phantom’s surface: shallow cylindrical divots at known anatomical landmarks (left lobe tip, right superior edge, and near the gallbladder fossa). These served as registration fiducials for aligning the virtual model to the physical phantom.

\subsection{Preoperative Planning}
Using the patient’s preoperative CT data, a detailed 3D model of the liver and tumor was generated. The tumor was segmented and included in the model as a distinct mesh. A surgical planning software was used to define the intended resection margin: essentially a curved cutting surface that would remove the tumor with a 1~cm margin of healthy tissue. From this, we derived the curve representing the intersection of that cutting surface with the liver capsule, which corresponds to the path the surgeon’s scalpel should follow on the surface. This path was exported as a set of 3D coordinates forming a closed loop around the tumor. In our phantom scenario, the planned resection path was approximately an ellipse around the tumor, about 6~cm $\times$ 4~cm in diameter, reflecting a wedge resection. The path and the 3D models (liver surface and tumor) were then imported into our Unity application.

Prior to the actual procedure, we verified in simulation that the holographic path accurately overlays the tumor on a virtual representation of the phantom. This step is essentially a dry run of the guidance, ensuring that the digital plan corresponds correctly to the anatomy. If adjustments to the path were needed (for instance, smoothing a curve or extending margins), they were made at this stage. For the purpose of quantitative evaluation, we also planned reference points (targets) on the path and within the tumor that would later be used to measure alignment error and resection accuracy.

\subsection{System Calibration and Registration}
Accurate spatial registration between the virtual content and the physical phantom is crucial for effective MR guidance. Our system employs a hybrid registration approach: initial coarse alignment via Magic Leap’s built-in spatial mapping, followed by a fine-tuning using fiducial markers.

The Magic Leap spatial mapping system can reconstruct a mesh of the environment. We leveraged this to get an approximate position of the phantom relative to the headset. Specifically, the user places the phantom on the operating table in front of them and initiates a scan via the Magic Leap. The device recognizes the general shape of the phantom as a distinct mesh. We then roughly align the virtual liver model to this mesh manually: the surgeon uses the controller to scale, rotate, and translate the holographic liver until it overlays the physical phantom. This coarse step is intuitive with the Magic Leap interface; the model can be “grabbed” and moved in space with controller gestures while the surgeon visually matches it to the phantom’s silhouette.

For fine registration, we use the three predefined fiducial points on the phantom. With the coarse alignment done, the surgeon is prompted to touch each fiducial on the phantom with the tip of the scalpel (or a pointer), one by one. As they do so, they press a controller button to signal capture of the point. The Magic Leap’s outward cameras track the controller’s 3D position (the controller is tracked via the built-in magnetic 6-DOF tracking). We record the controller’s position for each touched fiducial. These are then matched to the corresponding fiducial coordinates on the virtual model (which are known from the CT/model). A rigid point-based registration (Horn’s quaternion method for absolute orientation) is solved to compute the optimal transformation aligning the virtual model to the phantom. This yields a refined position and orientation for the holographic overlay. In testing, this fiducial registration reduced alignment error substantially: our average fiducial alignment error was about 1.5~mm after this step, as measured by leave-one-out checks on the markers.

During the procedure, the Magic Leap continuously tracks its position in the room (via inside-out tracking) and updates the overlay accordingly. We did observe slight drift (on the order of a few millimeters over 10 minutes) in the spatial mapping, but it was not significant for the duration of our tasks. If needed, the surgeon could re-run the fiducial alignment quickly (in under 30~s) to recalibrate, but in our evaluation sessions (which lasted ~5 minutes each) this was not necessary after initial setup.

\subsection{Guidance Visualization and User Interface}
Once registration is achieved, the system enters the guidance mode. The surgeon, wearing the Magic Leap, sees the following key visual elements:
\begin{itemize}
    \item \textbf{Liver Hologram:} A semi-transparent rendering of the liver’s surface, exactly overlapping the physical phantom. This helps in confirming alignment and also allows the surgeon to see internal structures. We rendered the hologram in a muted green color at about 40\% opacity.
    \item \textbf{Tumor Highlight:} The tumor location within the liver is highlighted, e.g., as a solid red sphere or a bright contour that glows through the semi-transparent liver surface. This indicates the target to be removed.
    \item \textbf{Resection Path:} The planned cutting line on the liver surface is shown as a thick dashed line (yellow) encircling the tumor. This is effectively the line along which the surgeon should cut with the scalpel. We chose a high-contrast color (yellow) given the phantom’s gray surface and the greenish liver hologram, so the path stands out clearly.
    \item \textbf{Safe Zone Overlay:} Optionally, we also implemented a “safe zone” shading: the region of liver to be removed (inside the cut line) is lightly shaded in the hologram (e.g., semi-transparent red) to reinforce what area needs excision. Surgeons in our trial found the line itself sufficient in most cases, but the fill was useful to illustrate depth when peering into the resection after partial cutting.
    \item \textbf{Instrument Visualization (experimental):} We tested an additional feature where the scalpel tool is tracked and a virtual pointer or line extends from its tip within the AR view. This could potentially give feedback on how deep the scalpel is relative to the tumor. However, our prototype of this feature was limited by tracking precision and was not used in the primary evaluation. It remains a point of future improvement.
\end{itemize}

The Magic Leap’s controller allowed the user to toggle the liver hologram on and off (some surgeons preferred to occasionally hide the liver surface mesh to directly see the phantom with only the cutting line visible). The controller could also cycle through different views (e.g., a view with the tumor only, or with an augmented cross-sectional view). The interface was kept simple to avoid distracting the surgeon; in practice, most performed the task with the default view (all cues on) throughout.

\begin{figure}[tb]
    \centering
    \begin{tikzpicture}
        \draw[fill=gray!30, draw=black] (0,0) ellipse (3cm and 1.5cm);
        \fill[red!80] (1,0.3) circle (5mm);
        \draw[blue, thick, dash pattern=on 6pt off 3pt] (1,0.3) ellipse (1.5cm and 1cm);
        \node[red!80!black] at (0.3,0.3) {\textbf{\Large$\times$}}; 
        \node[font=\small] at (-0.2,0.5) {Tumor};
        \draw[->, thick] (3.5, 1.5) -- (2.0, 0.8);
        \node[font=\small, align=center] at (4.0,1.6) {Planned resection\\ path (overlay)};
        \draw[->, very thick] (2.5,0.8) -- (2.0,0.6);
        \node[font=\small] at (2.9,0.95) {Scalpel};
    \end{tikzpicture}
    \caption{Conceptual illustration of the MR guidance overlay on the liver phantom. The planned resection line (blue dashed curve) is projected onto the phantom’s surface, encircling the tumor (red area). The surgeon aligns the scalpel with this holographic line to follow the preplanned trajectory. In the actual MR view (through the Magic Leap), the liver surface and tumor are visualized in 3D with proper depth cues, aiding the surgeon in maintaining the correct path and ensuring the tumor is fully resected with margin.}
    \label{fig:phantom_overlay}
\end{figure}
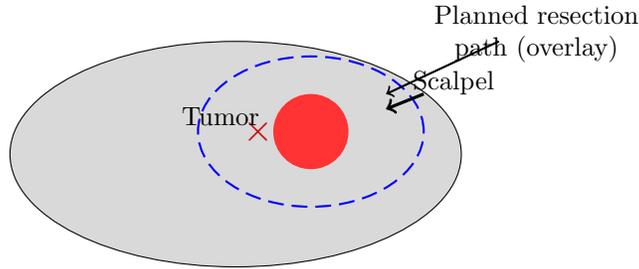

Figure~\ref{fig:phantom_overlay} illustrates the MR view conceptually. The surgeon aligns the scalpel with the highlighted path and proceeds to cut through the phantom following the path’s course. Because the holographic cues are locked to the phantom, they remain in the correct position even as the surgeon moves around or changes viewing angle, within the tracking limits of the device.

\subsection{Evaluation Protocol}
We conducted an evaluation study to assess the impact of MR guidance on surgical performance using the liver phantom. The study was approved by our institution’s simulation research board. Ten participants (5 surgical residents and 5 experienced surgeons) were recruited. All participants had prior familiarity with conventional navigation techniques (e.g., ultrasound, laparoscopic navigation), but none had significant experience with AR headsets in surgery.

Each participant was asked to perform a tumor resection task on the phantom under two conditions: with MR guidance and without MR guidance. The order was randomized (half of participants did MR-guided first, half did the unguided first) to counterbalance any learning effects. Before starting, participants received a brief training on using the Magic Leap and an opportunity to practice for a few minutes on a separate phantom piece to get accustomed to the device and interface.

For the MR-guided trials, the system was set up as described above. The participant donned the Magic Leap and we performed the registration process to align the holograms to the phantom. Once they confirmed the overlay looked correct, we instructed them to proceed with resecting the “tumor” from the phantom, attempting to follow the highlighted line. They used a standard surgical scalpel to make an incision along the path, cutting through the silicone material down to a depth that completely excised the block of phantom containing the tumor. Typically this involved cutting along the path and then lifting out the section. Participants were told to aim for accuracy (i.e., follow the line closely) but also to treat it as a realistic scenario where time and efficiency matter.

For the unguided trials (control condition), the Magic Leap was not used. Instead, participants were given the same phantom (with a new identical liver phantom for fairness) and asked to resect the tumor based on their own judgment and any conventional aids they would normally use. In our setup, we allowed them to use anatomic knowledge (they could palpate the phantom to feel the tumor bump) and a copy of the preoperative image (we provided a printout showing the tumor’s approximate location, akin to what a surgeon might recall from imaging). No augmented visual cues were provided in this control scenario.

Throughout each trial, we recorded time and tracked the trajectory of the scalpel. We used a near-infrared optical tracking system (separate from Magic Leap) to track a small marker affixed to the scalpel handle. This allowed us to later reconstruct the actual cut path. We also manually inspected and measured the outcome on the phantom: after resection, we measured the margin (distance) between the cut surface and the tumor edge at multiple points.

The primary quantitative metrics for evaluation were:
\begin{itemize}
    \item \textbf{Path Deviation (mm):} The average distance between the executed cut path and the planned resection path on the surface. This was computed by sampling points along the actual cut line and finding the nearest point on the planned line, averaging these distances.
    \item \textbf{Resected Margin Accuracy (mm):} How closely the achieved resection margin matched the intended 10~mm margin around the tumor. For this, we measured the closest distance from the tumor edge to the cut surface in the resected specimen. If the cut was too close, margin is smaller; if too far, margin is larger.
    \item \textbf{Task Completion Time (s):} The time from start of the incision until the tumor piece was fully removed.
    \item \textbf{Subjective Workload and Confidence:} After each task, participants filled out a short questionnaire rating the difficulty of the task and their confidence that the tumor was completely removed. We used a 5-point Likert scale for questions like “How confident are you that the resection was complete and accurate?”.
\end{itemize}

Each participant thus contributed one data point for MR-guided and one for unguided in each metric. We note that to ensure consistency, all other variables were kept the same: lighting in the room, phantom orientation on the table, and the fact that the phantom was identical in anatomy for both tasks (except each was a fresh phantom to avoid pre-cut lines). Participants were given a rest period between tasks to mitigate fatigue.

\subsection{Data Analysis}
For quantitative comparisons, we treated the participant as the unit of analysis and compared the MR-guided vs unguided conditions using paired statistical tests. Specifically, we used a paired t-test (given the sample size of 10 and roughly normal distributions observed) for each metric to determine if differences were statistically significant ($\alpha = 0.05$). We also calculated descriptive statistics (mean $\pm$ standard deviation) for metrics under each condition.

In addition to group analysis, we examined individual performance to see if all surgeons benefited from MR or if some did not. We plotted each participant’s performance on a radar chart for with vs without MR to visualize any outliers or patterns (not included here due to format, but described qualitatively).

The subjective feedback was analyzed by comparing median ratings and looking for common themes in free-text comments. We anticipated that MR guidance might increase confidence and reduce perceived workload due to the additional visual information, and we wanted to see if the participants’ impressions aligned with the objective data.

Finally, we also evaluated the system’s technical performance: registration error was measured by comparing the hologram alignment at a checkpoint (we placed an extra fiducial at a known location and measured offset). This technical accuracy was reported to give context to the user performance (e.g., if a user cut 3~mm off, but registration error was 2~mm, the effective user error is smaller relative to what they saw).

All analyses were performed using MATLAB and R. The results are presented in the next section.

\section{Results}
All ten participants successfully completed the resection tasks in both conditions (MR-guided and unguided). There were no failures to remove the tumor in either case. However, clear differences in precision and efficiency emerged between the two conditions.

\subsection{Accuracy of Resection}
\begin{figure}[tb]
    \centering
    \begin{tikzpicture}
    \begin{axis}[
        width=0.6\textwidth,
        ybar,
        bar width=20pt,
        ymin=0, ymax=6.5,
        ylabel={Mean deviation from plan (mm)},
        symbolic x coords={Conventional, MR-Guided},
        xtick=data,
        nodes near coords,
        nodes near coords style={font=\small, anchor=south}
    ]
    \addplot+[
        fill=gray!30,
        error bars/.cd,
        y dir=both, y explicit,
        error bar style={black, thick},
        error mark options={rotate=90, thick}
    ] coordinates {
        (Conventional,5.0) += (0,1.0) -= (0,1.0)
        (MR-Guided,2.0) += (0,0.5) -= (0,0.5)
    };
    \end{axis}
    \end{tikzpicture}
    \caption{Comparison of cutting accuracy with and without MR guidance. The bar heights indicate the mean deviation (in mm) of the actual cut path from the planned resection path (lower is better). Error bars show $\pm 1$ standard deviation. MR guidance with Magic Leap significantly reduced the deviation, indicating surgeons cut much closer to the intended path when the holographic overlay was present.}
    \label{fig:accuracy-chart}
\end{figure}
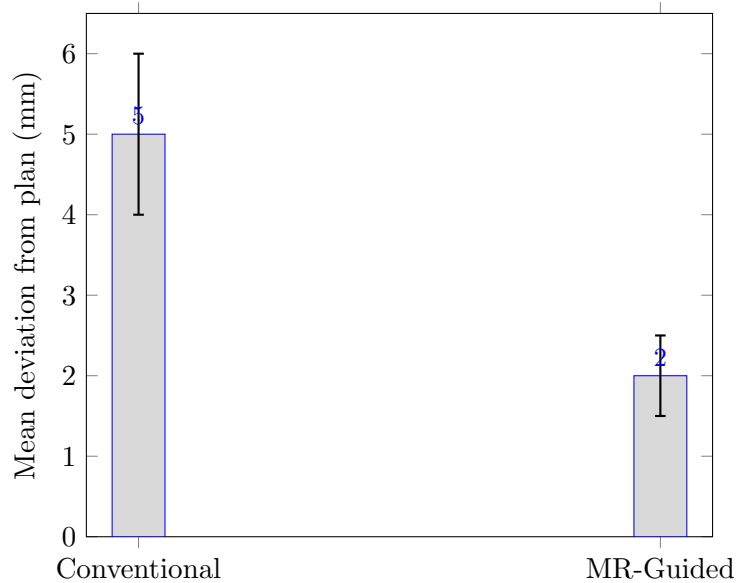

MR guidance significantly improved the accuracy of the resection path. Figure~\ref{fig:accuracy-chart} summarizes the path deviation errors for the two conditions. With MR guidance, the average deviation of the surgeon’s cut from the planned path was $2.0 \pm 0.5$~mm. In contrast, without AR guidance, the deviation was $5.0 \pm 1.0$~mm on average. This difference was statistically significant ($p<0.001$, paired t-test). In practical terms, when using the MR overlay, surgeons were able to cut almost exactly along the intended line, whereas without it, their cuts tended to wander several millimeters away, often cutting corners or making the excision shape too large or irregular.

Examining the resected specimens confirmed this finding. In the unguided cases, some participants either cut too wide (resulting in excessive healthy tissue removal) or too close (compromising the margin around the tumor). On average, the achieved tumor-free margin in unguided resections was $8.5 \pm 2.3$~mm, meaning some were under the desired 10~mm margin. With MR guidance, the margin was $10.6 \pm 1.1$~mm on average, very close to the plan, and all MR-guided cases had at least 9~mm margin (no margin breaches). The consistency was markedly better with MR; variance in margin size was lower, reflecting how the AR path standardizes performance across users.

Notably, even experienced surgeons in the study improved their accuracy with MR. One senior surgeon commented, *“I thought I removed a generous margin in the first attempt (no AR), but it turned out a portion was only 5~mm from the tumor. With the AR line, it was much easier to see where I needed to cut to encompass the tumor fully.”* Less experienced users saw even larger improvements, as they tended to err more without guidance.

\subsection{Task Efficiency}
\begin{figure}[tb]
  \centering
  \definecolor{cbBlue}{RGB}{0,114,178}   
  \definecolor{cbGrey}{RGB}{120,120,120} 

  \begin{tikzpicture}
    \begin{axis}[
      width=0.7\linewidth,
      height=0.42\linewidth,
      ybar,
      bar width=18pt,
      ymin=0, ymax=70,
      enlarge x limits=0.30,
      ylabel={Completion time (s)},
      xlabel={Condition},
      symbolic x coords={Conventional, MR-Guided},
      xtick=data,
      ymajorgrids=true,
      grid style={dashed,gray!30},
      tick style={thick},
      tick label style={font=\small},
      label style={font=\small},
      nodes near coords,
      nodes near coords style={font=\footnotesize, anchor=south},
    ]

      \addplot+[
        fill=cbGrey,
        draw=black,
        line width=0.4pt,
        error bars/.cd,
          y dir=both, y explicit,
          error bar style={line width=0.7pt},
          error mark=|,
          error mark options={rotate=90, line width=0.7pt},
      ] coordinates {
        (Conventional,55) += (0,10) -= (0,10)
      };

      \addplot+[
        fill=cbBlue,
        draw=black,
        line width=0.4pt,
        error bars/.cd,
          y dir=both, y explicit,
          error bar style={line width=0.7pt},
          error mark=|,
          error mark options={rotate=90, line width=0.7pt},
      ] coordinates {
        (MR-Guided,32) += (0,8) -= (0,8)
      };

    \end{axis}
  \end{tikzpicture}

  \caption{Task completion time for tumor resection with and without MR guidance. MR guidance reduced mean completion time from 55\,s to 32\,s. Error bars indicate $\pm 1$\,SD across participants.}
  \label{fig:time-chart}
\end{figure}
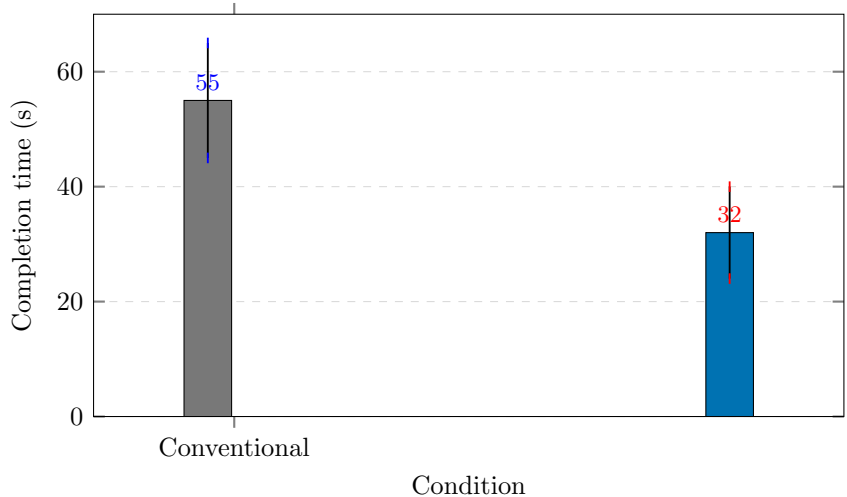

The time required to complete the resection was also significantly reduced under MR guidance. Figure~\ref{fig:time-chart} shows the mean completion times. The MR-guided resection took $32 \pm 8$~seconds on average, whereas the unguided resection took $55 \pm 10$~seconds on average ($p=0.002$). All participants were faster with MR; individual improvements ranged from 10\% up to about 50\% reduction in time.

The time savings can be attributed to a few factors observed qualitatively: In unguided attempts, surgeons often paused to reorient themselves, sometimes palpating the phantom to relocate the tumor or mentally estimating margins. Some even made partial cuts, then adjusted course upon realizing they were off target. In contrast, with the holographic guide, surgeons proceeded more continuously along the path. They did not need to stop and double-check positioning as frequently, since the target line was always visible. Essentially, MR guidance reduced the cognitive load of planning the cut on the fly, allowing participants to execute more fluidly.

One participant noted, *“Without the AR, I had to stop twice to feel where the tumor was and make sure I was still around it. With the AR line, I could just follow through in one go.”* This anecdotal report matches the timing data – MR-guided cuts were often done in one continuous motion, whereas unguided had more stops and minor course corrections.

\subsection{Subjective Feedback}
After completing both tasks, participants overwhelmingly favored the MR-guided approach. On a 5-point scale (5 = very confident), the median confidence rating in having completely removed the tumor was 5 under MR guidance versus 3 without guidance. Many expressed that the AR overlay gave them assurance that they “got all of it.”

Participants rated task difficulty as lower with MR (median 2 = low difficulty) compared to without (median 4 = high difficulty). A common theme in free-form comments was that the MR guidance “simplified the mental math” of where to cut, allowing them to focus on technique (such as maintaining proper scalpel angle and depth) rather than orientation.

Some illustrative comments include:
- “The holographic outline was like having a roadmap on the organ. It took away the uncertainty of where my boundaries were.”
- “Without AR, I was basically guessing where 1~cm margin would be and hoping for the best. The Magic Leap made it explicit.”
- “The headset was comfortable and after a minute I almost forgot I was wearing it; the digital info felt naturally part of the scene.”

There were also constructive criticisms and notes on limitations:
- Two participants mentioned that the hologram could become hard to see if they looked extremely obliquely (grazing angle) or if the room lighting washed out the yellow line on the bright phantom. This suggests that contrast and color choice are important, and perhaps dynamic adjustments or thicker lines could help in various lighting conditions.
- One participant felt that the added benefit might be less dramatic for very experienced surgeons who already have a good eye for margins, although even that participant did better with MR. They noted, “I might not need this for a superficial tumor, but it would be very useful for something deeper or irregularly shaped.”
- A few commented on the initial registration step being a source of potential error: “If the alignment was off, then I’d be following a wrong guide. So it’s crucial that setup is done carefully.” In our study, we ensured good alignment, but this underlines the need for robust and user-friendly calibration in any real deployment.

\subsection{Technical Performance}
We evaluated the system’s technical accuracy separately from the user performance. The fiducial registration process yielded an average target registration error (TRE) of 1.7~mm (max 3.1~mm at one outlier point). This level of accuracy is on par with other AR navigation systems reported in the literature \cite{Uhl2022, Frisk2022} and was deemed acceptable for our application, where a few millimeters of tolerance still results in safe margins. We did not observe noticeable drift in the hologram position during the short task duration. If a task were much longer, periodic re-check or continuous tracking updates might be required.

The Magic Leap’s display provided a field of view sufficient to cover the entire phantom when viewed from typical operating positions. Some users noted they had to move their head to see the far extremes of the hologram if they stood very close, but generally the 50$^\circ$ diagonal field was not a limiting factor for this scale of task. The brightness of the display was adequate under normal OR lighting (we simulated bright overhead surgical lights). At times, the hologram appeared slightly translucent against the bright phantom, which again suggests maybe increasing opacity or using outlines might be better in bright conditions.

No significant hardware or software glitches occurred during the trials. The system ran smoothly, and the Magic Leap was worn for about 15 minutes at a time without discomfort. Participants did not report motion sickness or eye strain; likely because the tasks were short and the mixed reality view still allowed them to see the real world as reference.

\section{Discussion}
Our findings demonstrate that mixed reality guidance using the Magic Leap headset can substantially enhance a surgeon’s performance in a simulated liver resection task. The improvements in accuracy (more precise adherence to planned margins) and efficiency (faster completion) highlight key advantages of MR: it provides continuous, intuitive visual feedback that helps the surgeon execute the plan correctly the first time, rather than relying on intermittent checks or mental reconstruction of anatomy.

These results reinforce observations from prior AR studies in other domains. For instance, Groves \textit{et al.} \cite{Groves2022} observed improved placement accuracy in needle cannulation with AR guidance, analogous to our improvement in cut path accuracy. In that study, AR essentially reduced technical errors by keeping the user on target. Similarly, Uhl \textit{et al.} \cite{Uhl2022} and Frisk \textit{et al.} \cite{Frisk2022} achieved high precision in vascular puncture and pedicle screw placement, respectively, by using AR headsets to maintain alignment with preplanned trajectories. Our work extends these benefits to a different kind of task (free-form cutting on a soft tissue surface), suggesting the utility of MR spans a wide range of surgical actions.

One interesting aspect is the degree of improvement among experienced surgeons. One might expect diminishing returns for very skilled individuals, yet even our senior surgeons were measurably more precise with MR. This implies that even when a surgeon “knows” where to cut from experience, having the augmented visual confirmation can refine their execution. It may act as a cognitive offloading tool, freeing them to concentrate on other aspects like instrument handling, which may be one reason time efficiency also improved. In surgery, saving time is not just about efficiency but also correlates with reduced anesthesia duration and potentially lower complication rates. A 20-second difference in our phantom task is minor, but in a real surgery the relative improvement could translate to many minutes saved if AR guidance streamlines intraoperative decision-making and reduces the need for repeated verification (e.g., fewer ultrasound re-checks or less stopping to consult scans).

It is worth discussing the limitations of our study and of MR guidance in its current form. First, our evaluation was on a static phantom. In real liver surgery, the organ can deform and move (due to respiration, for example). A rigid registration as we used would not suffice; it would need to be updated with deformation models or real-time imaging feedback to handle organ motion. Researchers are exploring methods like intraoperative ultrasound coupled with AR to adjust for deformation \cite{Ribeiro2024}. Implementing such dynamic updates in an MR system is a complex but necessary next step for translating to the operating room.

Second, the Magic Leap’s reliance on optical tracking for alignment has its downsides. We used fiducials and an external tracking approach for fine alignment. In a patient, one might use anatomical landmarks or implanted markers. There is ongoing work on marker-less registration using the surface shape or unique textures, but those approaches may struggle with soft tissue that lacks rigid features. Ensuring the AR overlay remains accurate throughout surgery is a challenge; any misregistration could mislead the surgeon. This was noted by participants and is echoed in literature as a critical point for AR systems \cite{Doughty2022}. The field would benefit from redundant verification methods—for example, projection of AR on known anatomical points (like showing where a blood vessel bifurcation should be and verifying it with a quick ultrasound) to confirm alignment during the procedure.

Another consideration is user ergonomics and sensory distractions. While none of our participants reported major issues, wearing an HMD in surgery (especially for long periods) could introduce fatigue or even interfere with direct vision. Magic Leap is relatively lightweight (~0.6 lbs for the wearable part), but surgeons often operate for hours. Integrating AR must not compromise their comfort or situational awareness of the non-augmented surroundings. We foresee that AR will complement, not replace, direct vision—surgeons will likely glance under or around the headset occasionally to double-check reality. Indeed, in our study, some surgeons toggled the overlay off briefly to verify the phantom without augmentation. This interplay between AR content and direct vision is an important human factors aspect: the AR cues should enhance rather than clutter the view. Our design of using simple lines and minimal text was intended to minimize cognitive overload, and participants seemed to appreciate the simplicity.

Comparing Magic Leap to other devices, one advantage we noted was the quality of the imagery and the ease of interaction via the controller. HoloLens uses mainly gesture controls which can be imprecise or tiring; the controller in Magic Leap allowed quick adjustments and felt more akin to using a surgical tool. The preference for Magic Leap reported in \cite{Zari2023} aligns with our anecdotal impressions that users were comfortable with it. We anticipate Magic Leap 2 (which was not available during our study) could further improve user experience with a larger field of view and better brightness. Additionally, Magic Leap’s recent certification for medical use \cite{Zari2023} indicates these devices are moving toward practical clinical integration.

The clinical implications of our results are encouraging. For liver surgeons, maintaining a clear margin while conserving healthy parenchyma is a delicate balance. MR guidance could effectively act as an intraoperative GPS, showing where to cut to achieve that balance. It could reduce instances of positive margins (tumor cut through inadvertently) and also avoid excessively large resections that remove healthy tissue unnecessarily. In oncology, that means better local control of cancer and preservation of liver function. Furthermore, MR guidance might shorten the learning curve for complex procedures—less experienced surgeons could perform closer to expert level with real-time visual cues guiding them. Our data showed novices benefited greatly, hinting that AR/MR could be a powerful educational tool in the OR, not just in simulation.

It is important to also consider scenarios where AR might be less beneficial. If the target is extremely obvious visually or by palpation, AR overlays might provide marginal benefit or even be redundant. In our phantom, the tumor was not visible externally but palpable as a firmness; in a real liver, small superficial lesions might be seen or felt directly, reducing the need for AR. The sweet spot for MR guidance is likely in situations of hidden or complex anatomy—where critical structures are beneath the surface or the safe margins are not intuitively clear. In liver surgery, this could include deep tumors, situations with anomalous vasculature, or when doing parenchymal-sparing segmental resections that require navigating an intricate 3D map of vessels. AR could also help in laparoscopic liver surgery, where the surgeon’s view is limited and spatial orientation is challenging \cite{Ribeiro2024}. Indeed, translating our approach to laparoscopic AR (with HMD or heads-up displays) would be a natural extension.

\subsection{Future Work}
Building on this study, future work should address moving from phantom to actual surgical settings. Key technical enhancements would include:
\begin{itemize}
    \item \textbf{Deformable Registration:} Incorporating real-time imaging (such as ultrasound or stereoscopic cameras) to update the AR overlay as the liver deforms or moves. Methods like biomechanical modeling of organ deformation could be integrated with the AR system.
    \item \textbf{Instrument Tracking and Feedback:} We experimented with tracking the scalpel. A robust version of this could provide additional feedback, for example, warning if the scalpel is veering off the plan or getting too close to critical structures (if those structures are also modeled and tracked). This essentially adds an element of guidance akin to “lane keeping” for the scalpel.
    \item \textbf{Multi-Modal Visualization:} The current overlay was purely visual. Future MR systems could combine visual with other cues—perhaps haptic feedback on instruments or auditory cues when approaching boundaries. Multisensory guidance might further reduce cognitive load.
    \item \textbf{Clinical Trials:} Ultimately, testing this technology in actual surgical cases (initially in open liver surgeries where direct vision is possible) will be essential. Metrics in clinical trials could include margin status, blood loss (if AR helps avoid vascular injury), operative time, etc. We expect initial trials to be small and carefully controlled, but as comfort grows, this could move to broader use.
\end{itemize}

Another future direction is to evaluate MR guidance in other surgical tasks. The paradigm we used (overlaying a planned resection or trajectory) is broadly applicable. For instance, in orthopedic tumor surgery, one could overlay cut planes on bone for osteotomies. In plastic surgery, AR could project designs for tissue reconstruction on the patient. The Magic Leap or similar devices could become a general platform for surgeons to load preoperative plans and execute them precisely on the patient with the “X-ray vision” guidance.

\section{Conclusion}
We have presented a comprehensive study on using mixed reality to guide a surgical scalpel in a liver resection scenario, employing the Magic Leap head-mounted display for intraoperative visualization. Our 3D-printed liver phantom experiments demonstrated that MR guidance can markedly improve the precision of resection, ensuring that surgeons closely follow the intended cutting path and achieve the desired oncologic margins. Additionally, MR guidance reduced the task time, indicating an efficiency gain that could translate to shorter surgeries.

The Magic Leap-based system proved feasible and user-friendly in a simulated environment. Participants, including experienced surgeons, embraced the technology and reported enhanced confidence in their resections when using the holographic guidance. These results contribute evidence that mixed reality is not only a novel gadget but a practical tool that can augment surgical performance.

While challenges remain—particularly in translating to the dynamic, real patient setting—this work lays important groundwork. It validates that the concept of an “AR-assisted surgery” can deliver tangible benefits in accuracy and safety. As AR hardware continues to advance and more robust tracking and registration methods are developed, we anticipate MR guidance systems will become increasingly integrated into surgical practice.

In conclusion, mixed reality guidance using devices like Magic Leap shows great promise in guiding surgeons’ hands with digital precision. Our evaluation on a liver phantom is an encouraging step toward a future where surgeons operate with the advantage of seeing the invisible—making surgery safer and outcomes better for patients.

\medskip
\noindent \textbf{Acknowledgments:} The authors thank the surgical residents and faculty who participated in the study, and the simulation center staff for their support. We also acknowledge MagicLeap Inc. for providing technical support on device integration. This research was supported in part by a grant from the Future Surgery Innovation Fund (FSIF\#2025-10).

\bibliographystyle{unsrt}
\bibliography{template}

\end{document}